# MODELLING SPATIAL PATTERNS OF ECONOMIC ACTIVITY IN THE NETHERLANDS

Yang, Jung-Hun
PhD Student
Faculty of Geosciences, Utrecht University
P.O.Box 80115, 3508 TC Utrecht
The Netherlands
Phone: (+31 30) 253 2918
Fax: (+31 30) 253 0604
E-mail:j.yang@geo.uu.nl

Ettema, Dick
Associate Professor
Faculty of Geosciences, Utrecht University
P.O.Box 80115, 3508 TC Utrecht
The Netherlands
Phone: (+31 30) 253 2918
Fax: (+31 30) 253 0604
E-mail: d.ettema@geo.uu.nl

Frenken, Koen
Associate Professor
Faculty of Geosciences, Utrecht University
P.O.Box 80115, 3508 TC Utrecht
The Netherlands
Phone: (+31 30) 253 2466
Fax: (+31 30) 253 0604
E-mail: k.frenken@geo.uu.nl

Van Oort, Frank
Professor
Faculty of Geosciences, Utrecht University
P.O.Box 80115, 3508 TC Utrecht
The Netherlands
Phone: (+31 30) 253 2230
Fax: (+31 30) 253 2746
E-mail: f.vanoort@geo.uu.nl

Visser, Evert-Jan
Department of Spatial Economic Policy,
Directorate General Enterprise and Innovation, Ministry of Economic Affairs
Bezuidenhoutseweg 20,
P.O.Box 20101 2500 EC Den Haag The Netherlands
Phone: (+31 30) 253 2620
Fax: (+31 70) 379 6095
E-mail: ejvisser@minez.nl

**Abstract:** Understanding how spatial configurations of economic activity emerge is important when formulating spatial planning and economic policy. Not only micro-simulation and agent-based model such as UrbanSim, ILUMAS and SIMFIRMS, but also Simon's model of hierarchical concentration have widely applied, for this purpose. These models, however, have limitations with respect to simulating structural changes in spatial economic systems and the impact of proximity. The present paper proposes a model of firm development that is based on behavioural rules such as growth, closure, spin-off and relocation. An important aspect of the model is that locational preferences of firms are based on agglomeration advantages, accessibility of markets and congestion, allowing for a proper description of concentration and deconcentration tendencies. By comparing the outcomes of the proposed model with real world data, we will calibrate the parameters and assess how well the model predicts existing spatial configurations and decide. The model is implemented as an agent-based simulation model describing firm development in the Netherlands in 21 industrial sectors from 1930 to 2004.

# 1. INTRODUCTION

Over the past decades microsimulation models of land use and transport have gained increasing interest as policy support tools. Especially in transportation and in land-use modelling, efforts have been made to develop comprehensive systems that describe the behaviours of relevant agents such as households, individuals, firms, developers and authorities. With the exception of developers and authorities, the agents can take decisions regarding their locations as well as about personal travel and the transport of goods.

The central tenet of these so called integrated land use and transport (ILUT) models is that both types of decisions are interrelated. Locational decisions are (at least partly) made based on the accessibility of other people and institutions and the implications of the location for travel and transport. In return, travel and transport behaviour is determined by the spatial distribution of households and firms stemming from individual locational decisions. The relationship between locational decisions and travel/transport behaviour implies also a relationship between households/ individuals on the one hand and firms/institutions on the other hand. For instance, households will base their locational decision on the accessibility of jobs, services and goods offered by firms. Likewise, firms will take their accessibility for workers and clients into account in their location choice.

Given these interactions, it is obvious that insight into firms locational decisions, leading to spatial configurations of economic activities, is vital if we are to predict the transport and land use implications of policies in the next decades with any degree of realism. Such implications could involve transportation effects, social implications of changes in accessibility, economic viability or ecological effects. Given this importance, however, it is concluded that the modelling of the spatial distribution of economic activity takes place under some limiting assumptions.

A first type of models (UrbanSim, SimFirms, ILUMASS) describes the evolution of spatial economic systems as a stochastic process, in which events such as firm growth, firm relocation, spin offs and take place with a probability that is predominantly a function of firm characteristics. In UrbanSim, economic activity is represented in terms of individual jobs, which are taken from an independent economic forecasting model, and are exogenous to the model. The jobs are treated as independent entities (i.e. not organised in firms), which are distributed across grid cells. ILUMASS (Moeckel, 2005) applies a more elaborate economic component. In particular, it uses a synthetic database of firms, which may take decisions regarding relocation, growth and closure. In addition, new firms may emerge at a particular birth rate, which is specific per sector and dependent on general economic growth rates. One of the most elaborate micro-simulation models of firms' developed to date is SIMFIRMS (Van Wissen, 2000). This model distinguishes the same events as ILUMASS (birth, growth, (re-) location, closure) but uses more sophisticated behavioural rules, accounting for such factors as market stress, spin offs of existing firms, age effects and spatial inertia in the case of relocation. Market stress is related to the concept of carrying capacity, which, analogous to the ecological concept, indicates the maximum number of firms that an urban system can contain. Carrying capacity is operationalised as the difference between market supply and market capacity, which is based on aggregate input-output models. Thus, the measure is the outcome of aggregate conceptualisations, rather than on firms' perception of demand and supply. In general the micro-simulation approaches are especially insightful to study demographic processes. For instance, they suffice to describe what the distribution across sectors in a region will be given some initial setting and given birth rates, spin-off probabilities etc. An element that is much less developed in these models is the role of spatial proximity. The fact that firms cluster in order to achieve agglomeration advantages is not well represented. Structural changes in spatial economics structures (e.g. the emergence of new economic centres due to changes in industries) are not well represented.

A second type of models focuses on the emergence of hierarchies of concentrations (of firms or population) as a result of simple reproduction and migration rules. Simon (1955) shows that by assuming fixed reproduction rates and relocation probabilities, and assuming that larger concentrations attract more migrants than lower concentrations, a hierarchy of concentrations emerges that follows a power law distribution. Remarkably, such power law distributions match existing hierarchies in economic concentration (Frenken et al., 2007) and population concentrations (Pumain, 2006) very well. Although apparently these simple reproduction and migration rules touch upon general principles of spatial organisation, the theoretical underpinning of the models is somewhat cumbersome (Krugman, 1996). In their most basic form, models as suggested by Simon are non-spatial. That is to say, the relative position of a concentration (e.g. a city or a commercial area) to other concentrations does not matter, since locational preferences of migrants only depend on the size of the concentration and not on its surroundings. As a result, a big city on an isolated place would be equally attractive as an equally big city surrounded by other cities. This assumption is problematic since it ignores the impact of proximity. For instance, studies in evolutionary economics (Boschma et al., 2002) suggest that proximity to other firms matters for their productivity and innovative capacity, and that this proximity exceeds the purely local scale. In particular, regions play an important role in processes of economic innovation, where the size of a region differs between types of industries. Thus, although correctly reproducing the rank size distribution of existing economic and population concentrations, the Simon model falls short in describing the emergence of clusters of economic development on a regional level.

In an earlier paper (Yang et al., 2008), we proposed an alternative approach to modelling the spatial distribution of economic activity that accounts for the impact of spatial proximity on spatial configurations. In particular, our approach distinguished between three different proximity effects:

- Accessibility to markets. This implies the number of potential clients and the distance one needs to travel to access these clients.
- Agglomeration effects. These encompass the competitive advantage that firms may gain by interacting with other related firms. This advantage may arise from exchange of knowledge through formal or informal channels, from shared (and cheaper) use of common facilities and by providing a larger 'draagvlak' for supporting services.
- Congestion. This represents the negative effects of concentration, such as traffic congestion and a stronger competition for services and facilities, leading to higher prices.

It is important to note that these proximity effects may work out differently for different types of firms, leading to different spatial distributions. First, agglomeration and accessibility to market are type specific effects, whereas the congestion is a general effect. That is to say, given some spatial distribution of firms, a location will offer agglomeration advantages for one type of firm but not for another type of firm. For instance, the City of London will offer agglomeration advantages for financial services, but not for the automotive industry. Real estate prices and traffic jams are however the same for everyone. Second, different types of firms will value the various proximity effects differently, leading to different locational decisions.

Knowledge intensive firms (e.g. R&D or banking) require much exchange of knowledge and information, so that the agglomeration advantages of an urban setting outweigh the disadvantage of congestion and high prices. Less innovative firms, such as production firms, will place a higher emphasis on low real estate prices and low congestion levels, leading to more peripheral locations.

In earlier paper (Yang et al., 2008), a series of simulations were carried out for a stylised setting in which the relative weight and the spatial scale of the proximity effects were systematically changed. These simulations indicated that different spatial configurations (centralised, multiple subcentres, shattered) can arise for different combinations of weight and spatial scale of the different proximity effects. These findings suggest that the proposed framework provides a promising starting point for modelling how changes in spatial configuration of economic activity occur over time, which would be an important contribution to ILUT models used for policy assessment. However, more elaborate tests are needed to find out to what extent models using the proximity concept are capable of representing structural changes in economic activity. To that end, the current paper presents simulation results with the model in a concrete setting. In particular, the model is used to 'predict' the development of the spatial distribution of economic activity in different sectors of the Dutch economy between 1950 and 2004. Given that the spatial distribution of firms is known for the base year 1950 and projection year 2004, comparison of the predicted and actual situation in 2004 allows us to draw conclusions about the usefulness of the proposed model.

The paper is organised as follows. Sections 2 describes the mechanics of the model, i.e. the behavioural assumptions made with respect to demographic development and (re)location behaviour. The data used to run the model is described in detail in section 3. Section 4 focuses on implementation issues pertaining to the choice of model parameters and the software organisation. Section 5 describes the outcomes of the simulations. Section 6, finally, draws conclusions and charts avenues for further research.

## 2. MODEL OUTLINE

The model used for the simulations is a slightly modified version of the model described in Yang et al. (2008). Like the previous model, it consists of two types of processes: demographic processes, such as firm growth, closure and spin-off, which basically determine the total number of firms and their average size and locational decisions, which may affect the spatial distribution.

### 2.1 Demographic Events

A first process that has to be modeled is firm growth. In this respect, we assume that macro-economic trends, leading to more or less activity in a certain economic sector on first instance are accommodated by adjusting the size of firms. To represent this we assume that the size of a firm *f* from sector *s* in year *t+1* (expressed as the number of employees) is a function of the size in the previous year:

$$S_{fst+1} = S_{fst}(1+ \varepsilon_s + \varphi_{fst}) \qquad (1)$$

It is assumed that the size of the firm increases with an amount $\varepsilon_s$ which represents the average growth in the sector. However the term can also represent a decline in size, due to economic trends or automation. The term $\varphi_{fst}$ is a stochastic term representing the fact that individual firms may differ in growth speed due to difference in management and circumstances.

Firm closure is modelled as a stochastic process, where each firm *f* from sector *s* in year *t* has a probability $P^c_{fst}$ of closing down. The probability is given by parameter $\vartheta_s$, which is specific for each sector. The probability can reflect trends in certain sectors.

$$P^c_{fst} = \vartheta_s + \rho_{fst} \tag{2}$$

It is noted that the net change in the number of firms is a function of both closures and the emergence of new firms. The last event is in this model assumed to take place through spin-offs. This is defendable, since many new firms are started by individuals who first worked for firms in the same sector. We assume that the probability that a firm will produce a spin-off depends on the size of a firm, relative to some reference size that is specific for the sector:

$$P^s_{fst} = \frac{1}{1 + \exp((\alpha_s \cdot S_{fst} - \beta_s)/S^s_{fst})} \tag{3}$$

This function implies that the spin-off probability increases with size according to an S-shaped function, with the parameters $\alpha_s$ and $\beta_s$ determining at what point and how quickly the probability increases. $S^s_{fst}$ is the critical firm size that determines at what size spin-offs start to become frequent. Once a spin-off occurs, the question is how large the spin-off will be. In practice, the size of the spin-off can vary between a single employee to a whole new division. To represent this variability, we represent the size of the spin-off as $\sigma_{fst}$, which follows some distribution that needs to be determined empirically. Assuming that small spin-offs occur more frequently than large ones, a right-skewed distribution such as lognormal would be appropriate. Firm sizes after spin-off can then be determined as:

$$S_{fst+1} = S_{fst} - \sigma_{fst} \tag{4}$$

$$S^{spin-off}_{fst+1} = \sigma_{fst} \tag{5}$$

**2.2 (Re)location decisions**

Apart from demographic developments, firms may decide to relocate. The relocation may be prompted by demographic events, for instance growth, necessitating moving to a larger facility. To model relocation events, we distinguish between three probabilities:

$\lambda_1$     the probability of not relocating;

$\lambda_2$     the probability of moving to a location where firms of the same type s already are located;

$\lambda_3$     the probability of moving to a location where firms of the same type s do not

yet exist.

The probabilities $\lambda_n$ could principally be derived from empirical data. For the current study we will however use probabilities that have proven to work well in other studies ($\lambda_1$ =0.9 etc.)

Once a firm has attempted to relocate to another location, it is assumed that it will move to the location $l$ which gives the highest utility $U^l$. $U^l$ is defined as a function of the three proximity effects identified before: market potential (MP), agglomeration potential (AP) and congestion (CP). Each is defined as a potential accessibility function relative to other firms:

$$MP_i = \sum_{.} N_j e^{-\alpha_1 d_{ij}}, \quad AP_i^{type} = \sum_{.} N_j^{type} e^{-\alpha_2 d_{ij}}, \quad CP_i = \sum_{.} N_j e^{-\alpha_3 d_{ij}} \qquad (6)$$

$$U^l = \beta_1 MP_l + \beta_2 AP_l^{type} + \beta_3 CP_l \qquad (7)$$

Note that where MP and CP are defined relative to all firms, AP is defined relative to firms of the same type *s*. In addition, we assume that utility of a new location is inversely related to the distance to the old location, representing the preference to stay close to existing clients and connected firms. This leads to the following utility function:

$$U^l = \delta_s d_{ll'} + \sum_s (MP_{ls} + AP_{ls} + CP_{ls}) \qquad (8)$$

## 3. DATA

### 3.1 Available Firm Data

The simulation study is carried out for the Netherlands for the period 1950-2004. For both years detailed data on the level of individual firms is available regarding a) size b) sector and c) municipality. Table 2 shows the total numbers of firms and employees and average firm size, aggregated to the Dutch notional level for the two years. This table clearly illustrates some of the fundamental changes taking place in the Dutch economy in the past half century.

Table 1: The classification of Industrial Sectors (CBS: Statistics Netherlands)

| Name | New Class | Class 1950 | Class 2004 |
| --- | --- | --- | --- |
| Pottery / Glass | 1 | 01, 02 | 26 |
| Publishing / Printing | 2 | 03, 49 | 22 |
| Construction | 3 | 04 | 45 |
| Chemical industry | 4 | 05 | 24 |
| Wood industry | 5 | 06, 08 | 20, 36 |
| Clothing production | 6 | 07 | 18 |
| Leather / Rubber | 7 | 09 | 19, 25 |
| Mining / Oil / Gas | 8 | 10 | 10, 11, 14 23 |
| Create industry | 9 | 11 | 27-35, 50 |
| Paper industry | 10 | 14 | 21 |
| Textiles industry | 11 | 15 | 17 |
| Utility industry | 12 | 16 | 40, 41 |
| Food industry | 13 | 17 | 15, 16 |
| Retail | 14 | 40-42, 43 | 52 |
| Wholesale | 15 | 45 | 51 |
| Transport by air | 16 | 50 | 62 |
| Inland | 17 | 51, 52 | 60 |
| Water transport | 18 | 53 | 61 |
| Services for transport | 19 | 54 | 63, 71 |
| Communications | 20 | 55 | 64 |
| Catering | 21 | 56 | 55 |

Table 2: The number of firms and employees, average firm size in the Dutch

| Sector | | 1950 year | 2004 year |
|---|---|---|---|
| All | Firms | 406444 | 438252 |
| | Employees | 2490489 | 3616323 |
| | Firm Size | 17 | 19 |
| 1 | Firms | 2074 | 2137 |
| | Employees | 50692 | 33006 |
| | Firm Size | 24 | 15 |
| 2 | Firms | 9206 | 8274 |
| | Employees | 72521 | 85594 |
| | Firm Size | 8 | 10 |
| 3 | Firms | 41790 | 76996 |
| | Employees | 294211 | 473520 |
| | Firm Size | 7 | 6 |
| 4 | Firms | 3501 | 1118 |
| | Employees | 56992 | 75676 |
| | Firm Size | 16 | 68 |
| 5 | Firms | 9000 | 10386 |
| | Employees | 64564 | 161132 |
| | Firm Size | 7 | 16 |
| 6 | Firms | 29137 | 1627 |
| | Employees | 165154 | 6471 |
| | Firm Size | 6 | 4 |
| 7 | Firms | 11978 | 1673 |
| | Employees | 58951 | 37893 |
| | Firm Size | 5 | 23 |
| | Firms | 936 | 314 |
| 8  | Employees | 52703  | 10337  |
|    | Firm Size | 56     | 33     |
| 9  | Firms     | 39726  | 48082  |
|    | Employees | 426990 | 518150 |
|    | Firm Size | 11     | 11     |
| 10 | Firms     | 455    | 462    |
|    | Employees | 23579  | 21831  |
|    | Firm Size | 52     | 47     |
| 11 | Firms     | 1965   | 1460   |
|    | Employees | 130211 | 16272  |
|    | Firm Size | 66     | 11     |
| 12 | Firms     | 861    | 459    |
|    | Employees | 33895  | 28376  |
|    | Firm Size | 39     | 62     |
| 13 | Firms     | 24116  | 4918   |
|    | Employees | 235842 | 130678 |
|    | Firm Size | 10     | 27     |
| 14 | Firms     | 130285 | 120458 |
|    | Employees | 313460 | 708681 |
|    | Firm Size | 2      | 6      |
| 15 | Firms     | 38236  | 71408  |
|    | Employees | 180119 | 485200 |
|    | Firm Size | 5      | 7      |
| 16 | Firms     | 2      | 358    |
|    | Employees | 6606   | 23863  |
|    | Firm Size | 3303   | 67     |
| 17 | Firms     | 20930  | 14681  |
|    | Employees | 102569 | 206096 |
|    | Firm Size | 5      | 14     |
| 18 | Firms     | 10476  | 3871   |
|    | Employees | 69722  | 21956  |
|    | Firm Size | 7      | 6      |
| 19 | Firms     | 2984   | 16324  |
|    | Employees | 29052  | 146130 |
|    | Firm Size | 10     | 9      |
| 20 | Firms     | 2345   | 5893   |
|    | Employees | 44970  | 115279 |
|    | Firm Size | 19     | 20     |
| 21 | Firms     | 26441  | 45353  |
|    | Employees | 77686  | 310182 |
|    | Firm Size | 3      | 7      |

The spatial system used for this simulation consists of all municipalities as zones. However, since the municipal level is too coarse for calculating proximity effects, which are assumed to determine the evolution in the spatial distribution of economic activity, an underlying system was created consisting of 125x106m grid cells. A critical point is then how to allocate firms, for which the municipality is known, to grid cells. Since this model is a first step in the development of more advanced simulation methods, we chose to randomly assign firms to grid cells in their municipality. Thus, firms that are present in 1950 are assigned to the grid cell system according to the above procedure, to create the starting point of the simulation.

## 4. PARAMETER SETTINGS AND SIMULATION PROCEDURE

### 4.1 Deriving parameters

Having defined the base data of the simulation, it is necessary that the parameters in growth, spin-off and closure functions are determined as well as the parameters in the location choice functions. The parameters of the growth, spin-off and closure functions were determined by trial and error, aiming at accurately predicting the number of firms and the number of employees at the aggregate level. The idea behind this is that changes in the spatial distribution will arise mainly from locational decisions. Basically, the number firms is determined by:

$$\#Firms_{t+1} = \#Firm_t - Closures_t + SpinOffs_t \tag{9}$$

The number of employees is determined as:

$$\#Empl_{t+1} = \#Empl_t + GrowthPerFirm_t - Closures_t \tag{10}$$

Thus, applying equations 9 and 10 (using equations 1 to 3 to determine growth, closure and spin-off) to the set of firms for 54 consecutive years gives a predicted total number of firms and employees in 2004. By manually manipulating the parameters of equations 1 to 3, the following objective functions were minimised for each sector:

$$\xi_s = |\#Empl_{2004}^{predicted} - \#Empl_{2004}^{observed}| \tag{11}$$

$$\psi_s = |\#Firms_{2004}^{predicted} - \#Firms_{2004}^{observed}| \tag{12}$$

This resulted in the following set of parameters:

Table 3: The set of population parameters

| Sector | $\varepsilon_s$ | $\vartheta_s$ | $S_{fst}^s$ | $\alpha_s$ | $\beta_s$ |
|---|---|---|---|---|---|
| 1 | -0.00646 | 0.00010 | 24 | 3.1 | 0.1 |
| 2 | 0.00334 | 0.00120 | 8 | 4 | 0.1 |
| 3 | 0.01129 | 0.00900 | 7 | 2.7 | 0.1 |
| 4 | 0.00607 | 0.01500 | 16 | 4 | 0.1 |
| 5 | 0.02770 | 0.00100 | 7 | 2.8 | 0.1 |
| 6 | -0.01779 | 0.00700 | 6 | 5.3 | 0.1 |
| 7  | -0.00662 | 0.01500 | 5    | 5.5 | 0.1 |
| 8  | -0.01489 | 0.02000 | 56   | 6   | 0.1 |
| 9  |  0.00395 | 0.00852 | 11   | 2   | 0.1 |
| 10 | -0.00137 | 0.00700 | 52   | 4   | 0.1 |
| 11 | -0.01620 | 0.01000 | 66   | 6.5 | 0.1 |
| 12 | -0.00302 | 0.00400 | 39   | 6   | 0.1 |
| 13 | -0.00862 | 0.01000 | 10   | 5   | 0.1 |
| 14 |  0.02335 | 0.00100 | 2    | 6   | 0.1 |
| 15 |  0.03137 | 0.00100 | 5    | 2   | 0.1 |
| 16 |  0.04838 | 0.00010 | 3303 | 1   | 0.1 |
| 17 |  0.01869 | 0.01294 | 5    | 3.5 | 0.1 |
| 18 | -0.01269 | 0.02000 | 7    | 5.5 | 0.1 |
| 19 |  0.07463 | 0.00100 | 10   | 0.1 | 0.1 |
| 20 |  0.02895 | 0.00100 | 19   | 1   | 0.1 |
| 21 |  0.05542 | 0.01147 | 3    | 1   | 0.1 |

Figure 1 illustrates for the sectors 'manufacturing textiles', 'paper industry' and 'communications' how the trend in these sectors is represented by the growth/closure/spin-off functions.

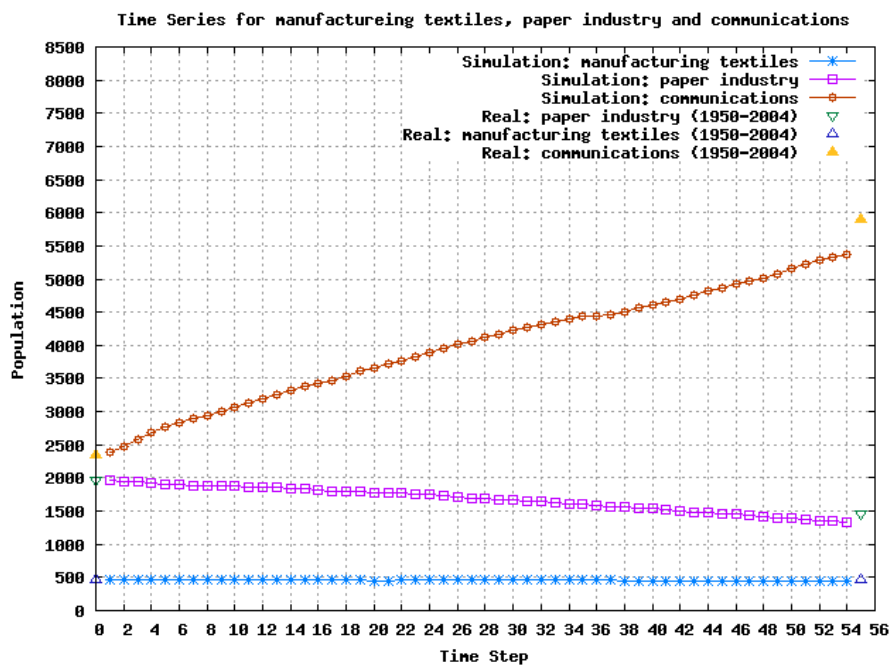

Figure 1 Time Series for manufacturing textile, paper industry and communications

The parameters of the location choice model are treated somewhat differently. Rather than fitting the parameters to reproduce the data, we intend to run different simulations with different sets of parameters, representing different locational preferences, to find out what parameters are critical when modelling spatial patterns of economic activity and which set of parameters produces typical patterns for each sector. In the current paper, however, time and paper length constraints prohibit an extensive testing of different sets of model parameters. Therefore the model is run with a single set f parameters, as described in Tables 4 and 5.

Table 4: The set of relocation parameters

|  | $\lambda_1$ | $\lambda_2$ | $\lambda_3$ |
|---|---|---|---|
| Probability | 0.9 | 0.09 | 0.01 |

Table 5: The set of potential parameters

| Sector | $\alpha_1$ | $\alpha_2$ | $\alpha_3$ | $\beta_1$ | $\beta_2$ | $\beta_3$ |
|---|---|---|---|---|---|---|
| 9 | 0.1 | 0.1 | 0.1 | 1 | 0.5 | -1 |
| 15 | 0.2 | 0.2 | 0.2 | 1 | 0.5 | -1 |
| 14 | 0.3 | 0.3 | 0.3 | 1 | 0.5 | -1 |
| 3 | 0.4 | 0.4 | 0.4 | 1 | 0.5 | -1 |
| 6 | 0.5 | 0.5 | 0.5 | 1 | 0.5 | -1 |
| others | 0.6 | 0.6 | 0.6 | 1 | 0.5 | -1 |

The simulation is run as an agent-based model. That is to say, each firm is updated each year in terms of number of employees, spin-offs, closure and relocation. Note that the relocation decision depends on the spatial distribution of other firms, but that each firm's evolution and locational decisions contribute to the spatial distribution. This leads to a repeated interaction between individual firms and the system level as illustrated in figure 2.

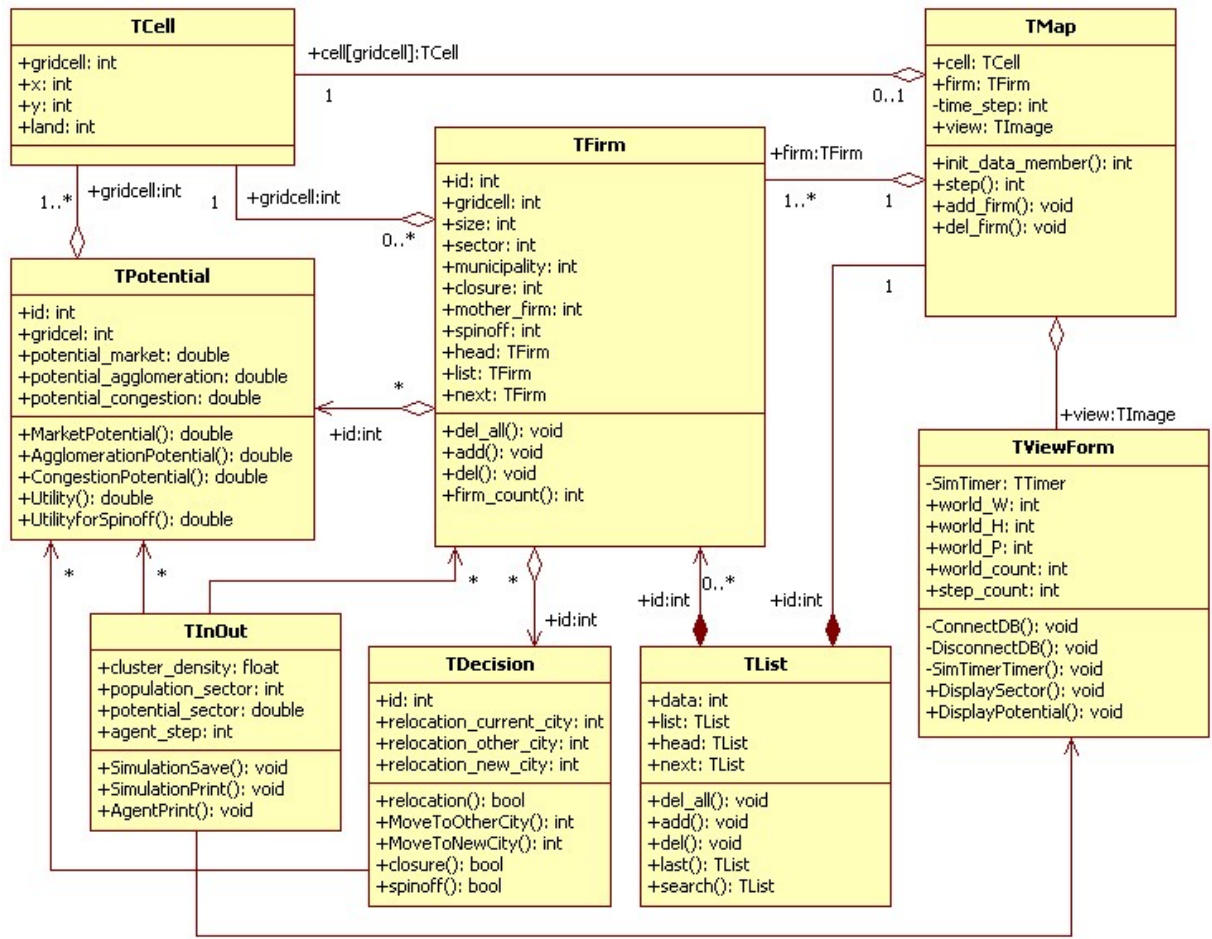

Figure 2 Design Model

## 5. SIMULATION RESULTS

The simulation predicts for a given year the number of firms and number of employees in each cell, for each sector. The simulation results are illustrated below for the sectors 'manufacturing industry' and 'wholesale'. For manufacturing industry (Figure 3) we note first of all that path dependency plays an important role. The 1950 pattern can be found back in subsequent years. This is logical since in areas with manufacturing firms, many firms will persist and create new firms through a spin-off process. These spin-offs are likely to be located near the mother firm. However, original centres of manufacturing may gain or loose in relative importance or disappear, as in the north of the country. Comparing the predicted outcome for 2004 to the actual pattern, we notice that the simulation outcomes bear some resemblance to reality to the extent that areas in which we simulate concentration of firms are also areas in which concentration is observed in reality. However, in reality, the number of areas with concentrations of manufacturing industries is larger, and also, the highest concentrations are observed in other areas. More testing of parameters will be required to achieve a more realistic result.

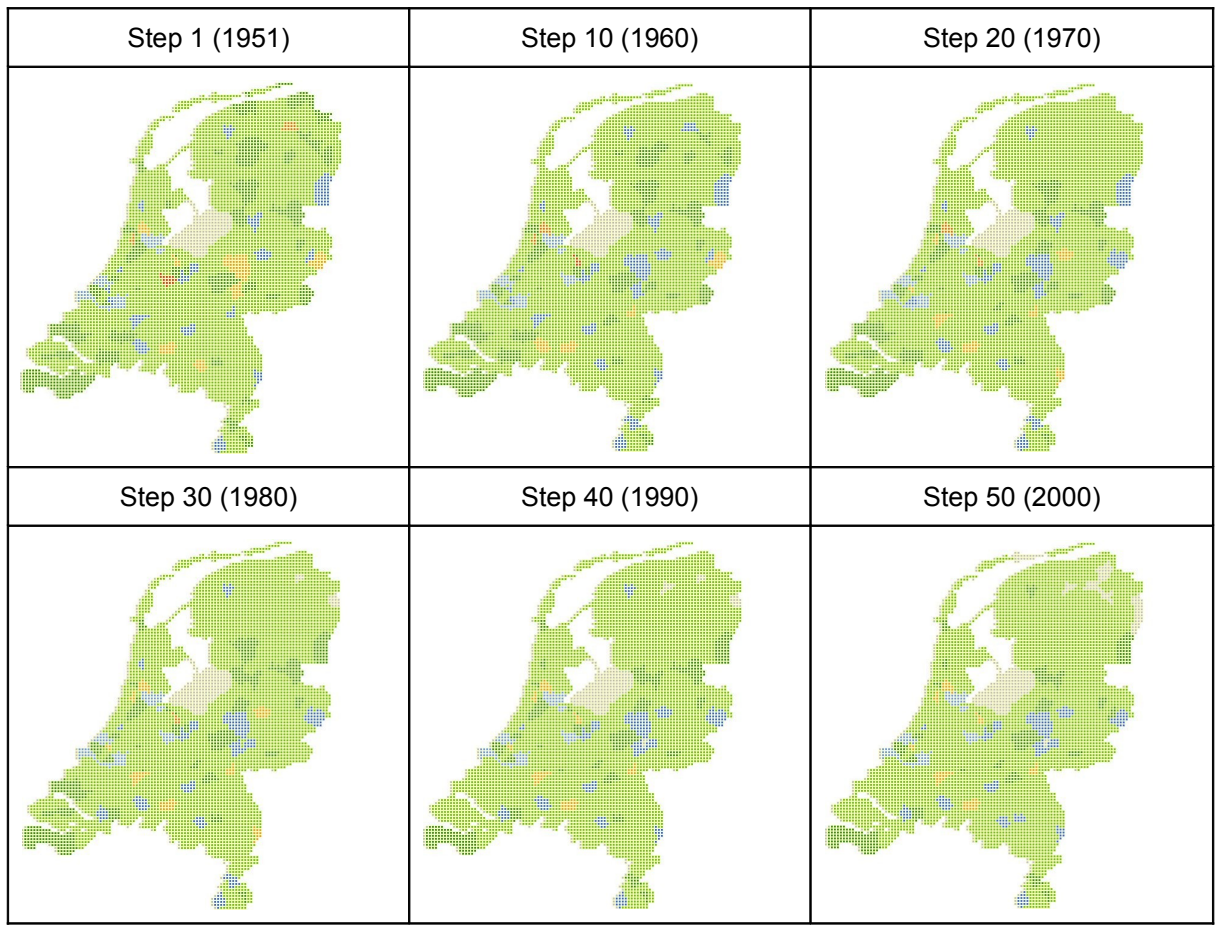

Figure 3 Spatial Patterns of Manufacturing Industry (unit: firms/municipality)

For wholesale (Figure 4) we observe a similar path dependency, with concentrations existing in 1950 persisting over longer periods. Also, the simulation suggests a trend of concentration toward fewer centres. In general, wholesale appears to be concentrated in larger concentrations of population and near borders. Comparing the simulations for 2004 to the actual spatial distributions, similar conclusions can be drawn as for manufacturing industry, suggesting that further improvement of parameters is needed.

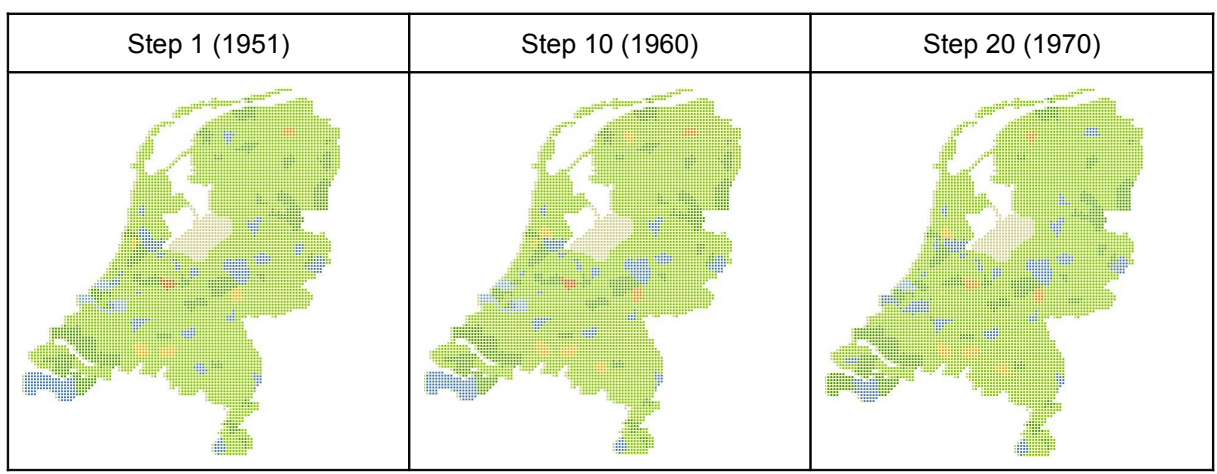

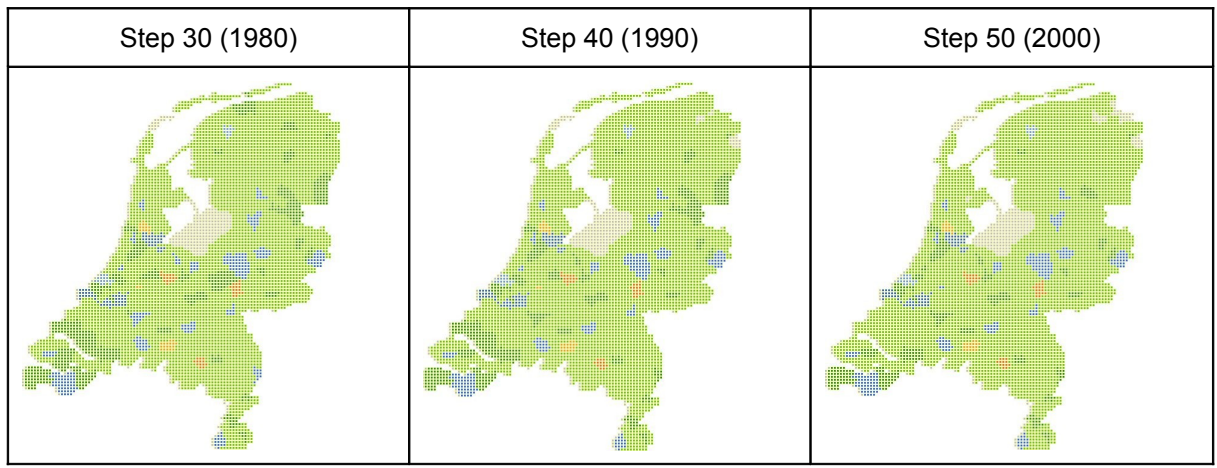

Figure 4 Spatial Patterns of Wholesale (unit: firms/municipality)

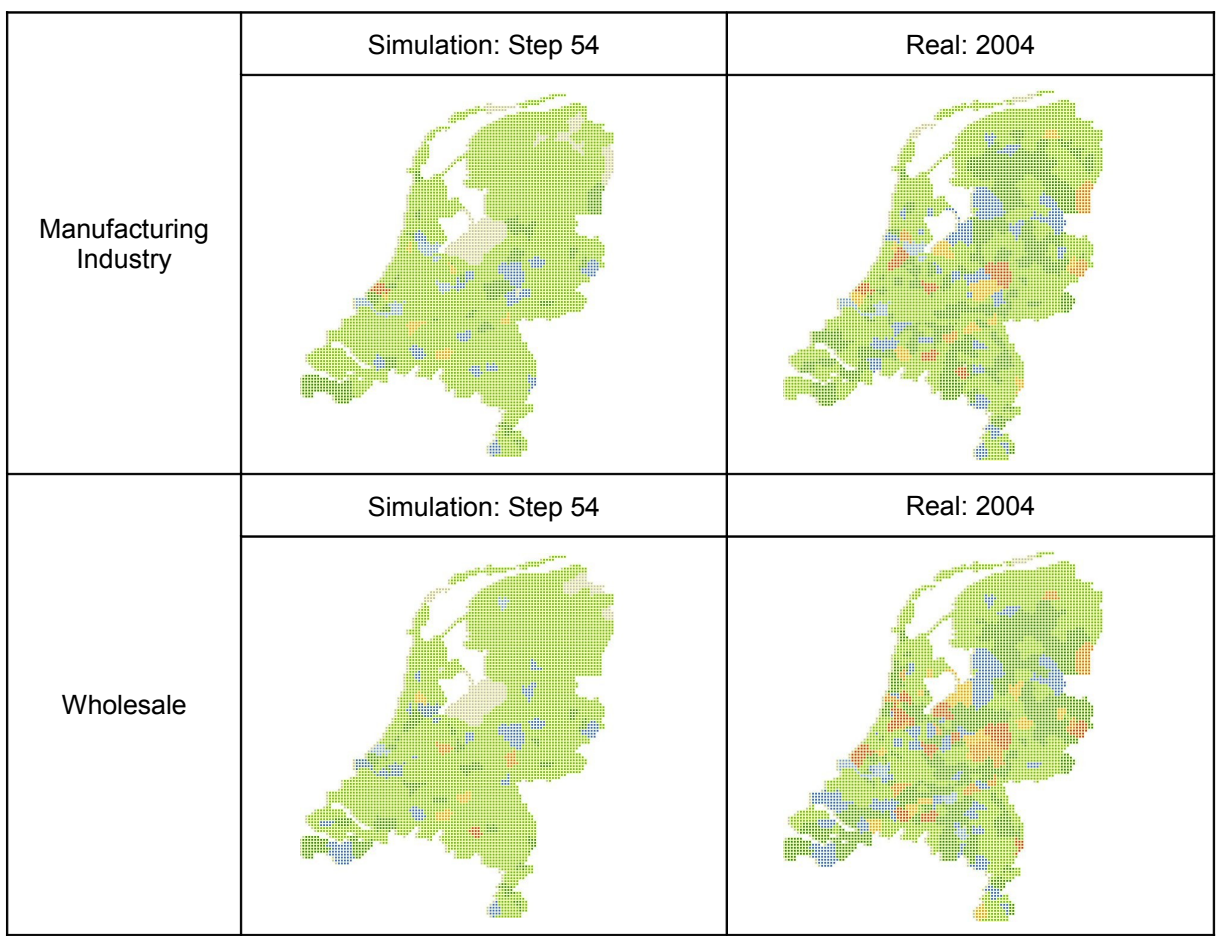

Figure 5 Comparison of Simulation and Actual Data (unit: firms/municipality)

## 6. CONCLUSIONS

This paper has proposed an approach to modeling spatial patterns of economic activity. Basically the model applies an agent-based approach, by modeling the evolution and spatial decisions of individual firms. To this end, functions of demographic events are specified as well as models of relocation probability and locational preference. To demonstrate the applicability of the model, it was applied to a set of 406,444 firms specified for the Netherlands in 1950. Functions of demographic processes were calibrated to match aggregate totals per sector, whereas the effects of different spatial preference functions are explored in scenario analyses. First application of the model suggests that path dependency, combined with general sectoral trends (growth/decline) plays an important role in predicting future patterns. This is justified to the extent that path dependency indeed plays an important role in reality. However, emergent developments, such as the growth of a sector in a new area, appear to be much harder to predict. In general more work needs to be done on calibrating the functions driving the model. One option would be to derive demographic models that are context-specific. That is to say, they should be able to respond to their environment in the probability of growth, decline and spin-off. Also, more work is needed on testing the effects of proximity on developments in particular sectors. This may be the key to understanding why some sectors are mostly path dependent and other are much more likely to display emergent behaviour. Finally, the linkage with population is currently lacking. Linking the model to available population statistics will improve the model, especially for sectors such as retail, for which proximity to population is a key factor.